\documentclass{aa}

\usepackage{graphics}

\newcommand{\be}{\begin{equation}} 
\newcommand{\ee}{\end{equation}} 
\newcommand{\ba}{\begin{eqnarray}} 
\newcommand{\ea}{\end{eqnarray}} 

\newcommand{\Msun}{\mbox{$M_{\odot}\;$}}

\begin{document}

   \thesaurus{08.14.1;   
              08.19.4;   
              08.16.6;   
              02.13.1    
             }

\title {General Relativistic Treatment of the
        Thermal, Magnetic and Rotational Evolution of Isolated Neutron Stars 
        with Crustal Magnetic Fields}

\author{Dany Page$^1$, Ulrich Geppert$^2$ \& Thomas Zannias$^3$}
\institute{1) Instituto de Astronom\'{\i}a, UNAM, 
              04510 Mexico D.F., Mexico; page@astroscu.unam.mx \\
           2) Astrophysikalisches Institut Potsdam,
              An der Sternwarte 16, 14482 Potsdam, Germany; urme@aip.de \\
           3) Instituto de F\'{\i}sica y Matem\'aticas, 
              Universidad Michoacana SNH, 
              58040 Morelia, M\'exico; zannias@ginette.ifm.umich.mx}
%
%
   \date{Received date ; accepted date}
%
 \titlerunning{Isolated Neutron Stars with Crustal Fields}
 \authorrunning{Page, Geppert, \& Zannias}
   \maketitle
%
\begin{abstract}
We investigate the thermal, magnetic and rotational evolution of isolated 
neutron 
stars assuming that the dipolar magnetic field is confined to the crust.
Our treatment, for the first time, uses a fully general relativistic formalism 
not only for the thermal but also for the magnetic part, and includes partial 
general relativistic effects in the rotational part.
Due to the fact that the combined evolution depends crucially 
upon the compactness of the star, three different equations of state
have been employed in the calculations.
In the absence of general relativistic effects, while upon  increasing  
compactness a
decrease of the crust thickness takes place leading  into an accelerating 
field decay,
the inclusion of general relativistic effects intend to 
``decelerate this acceleration''.
As a consequence we find that within the crustal field hypothesis,
a given equation of state is compatible with the observed periods $P$
and period derivative $\dot P$ provided the 
initial field strength and current location as well as the magnitude of the 
impurity content are  constrained appropriately.

Finally, we access the flexibility of the soft, medium and stiff
classes of equations of state as candidates in describing
the state of the matter in the neutron star interiors.
The comparison of our model calculations with observations, together with the 
consideration of independent information about neutron star evolution, 
suggests 
that a not too soft equation of state describes neutron star interiors and its 
cooling proceeds along the `standard' scenario.

\end{abstract}
%
\section{Introduction}

The question whether and, if so, how the magnetic field of 
isolated neutron stars (NSs) decays is a controversial
issue  and a subject 
of hot scientific debates.
The observed rotational periods $P$ and time derivative $\dot P$ 
of about  700 pulsars (PSRs)
(see PSR catalogue, Taylor et al. 1993\cite{TML93}) 
and studies of 
 their inferred surface magnetic field strength 
versus their active age ($\tau_a$) 
provide evidence that the magnetic 
fields of NSs is subject to 
decay. This evidence is rather strong
for the old population of PSRs, i.e. $\tau_a \gg 100$~Myrs, while
for the younger population
evidence for magnetic  field decay is much weaker. 
As investigated by population synthesis methods 
(Bhattacharya et al. 1992\cite{BWHV92}, 
Hartman et al. 1997\cite{HBWV97}), observations
and models are in harmony 
provided one accepts the hypothesis that NS magnetic field 
decays  very little during the first  $\tau_a \la 10$ Myrs. 
The NS magnetic field decay depends strongly on where is the field
located in the NS, what is its structure and strength. 
It is also related to the equation of state (EOS) of NS matter
and the conductive properties of dense matter which are moreover
affected by its thermal history too.
In case the field is of fossil origin, or alternatively has been 
generated via a  dynamo action during  the proto--NS phase  
(Thompson \& Duncan 1993\cite{TD93}), it is expected to thread most of the 
star. 
Early estimates of the electric conductivity of NS matter lead to the 
conclusion that magnetic field of NSs will not be  dissipated during a 
Hubble time (Baym, Pethick \& Pines 1969\cite{BPP69}).
However, more recent investigations regarding the conductive properties 
of nuclear matter in the presence of a strong B-field yield
conductivities leading to much shorter field decay times 
(see e.g. Haensel, Urpin \& Yakovlev 1990\cite{HUY90},
Goldreich \& Reissenegger 1992\cite{GR92}, Urpin \& Shalybkov 1995\cite{US95}, 
Shalybkov \& Urpin 1997\cite{SU97}).
Besides the conductive properties of nuclear matter, another distinct 
mechanism 
leading to a B-decay  process is based on the idea of  
magnetic flux expulsion from NS core driven by rotation or/and buoyancy 
(Muslimov \& Tsygan 1985\cite{MT85}; Srinivasan et al 1990\cite{SBMT90}).
Once the expelled field reaches the star's crust it subsequently
suffers Ohmic decay. 
Ding et al. (1993)\cite{DCC93} considered in detailed this  
process and have found typical decay times of  the order of $100$ Myrs. 
Recent investigations, which take into account the effect of the NS
crust onto the process of flux expulsion 
(Konenkov \& Geppert 2000\cite{KG00}) estimated even 
longer decay times for a field anchored in fluxoids in the 
superfluid NS core.

Although the issue of whether the field is penetrating
the entire star or part of it is an open one,
there exist, however, good reasons to 
believe that the NS magnetic field is maintained by 
currents in the crust. One possibility is the generation 
of the field via thermomagnetic effects during the first 
years of the NS life (Blandford et al. 1983\cite{BAH83}, 
Urpin, Levshakov \& Yakovlev 1986\cite{ULY86}, 
Wiebicke \& Geppert 1996\cite{WG96}). 
The most recent and detailed investigation of the magnetic and spin evolution 
of 
isolated NSs with a crustal dipolar field has been performed by 
Urpin \& Konenkov (1997\cite{UK97}, UK97 thereafter). 
They found a good agreement with observational data provided the EOS
is not too soft, the initial surface magnetic field strength lies in the range 
of $10^{12}$ to $3\times10^{13}$ G and is initially confined to densities of 
$10^{12}$ to $10^{13}$ g cm$^{-3}$ within the crust. 
Miralles, Urpin \& Konenkov (1998)\cite {MUK98} considered the effect of 
a crustal field decay onto the thermal evolution of a NS
and they have concluded
that  a considerable amount of 
heating takes place in the  the crust after about
$3$ to $10$ Myrs ,
a period during which  the NS has cooled down to $\la 10^5$ K. 
Consequently, Joule heating can maintain a warm ($\approx 5\cdot 10^4$ K) 
NS surface for a period of hundreds of Myrs.

All of the above described works ignore the curvature of spacetime. 
To date there exist no calculation treating self consistently the thermal and 
magnetic evolution of NSs which incorporate 
General Relativistic (GR) effects. 
As it has been argued elsewhere by the present
authors (Geppert, Page \& Zannias 2000\cite{GPZ00}, referred to as GPZ00 
hereafter), 
relativistic effects on the B-field evolution must be included
in detailed investigations.
A previous attempt to incorporate GR effects has been presented by
Sengupta (1997\cite{S97}, 1998\cite{S98}).
However this author failed to take into account the proper boundary 
condition associated with relativistic treatment and furthermore his
formalism applies only to a Schwarzschild geometry.
Moreover, he claims that the decay rate of the field is decreased by several 
orders of magnitude, a conclusion being in variance with the one 
obtain by GPZ00\cite{GPZ00} and the 
present work.

In the present paper we investigate in details and in a self 
consistent manner, GR effects at first on both: 
the thermal and magnetic evolution of NSs.
We consider three different EOSs and via 
numerical integration of Einstein's 
equations, neutron star models are constructed 
characterized by the compactness ratio varying in a large range.
We solve simultaneously the relevant evolution equations
and thus our approach naturally 
 reveals the mutual dependencies of EOS, mass--to--radius relation, 
initial field structure and strength, and thermal history of NSs. 
In particularly, as far as the cooling is concerned we consider two scenarios:
 the `standard'
slow neutrino emission scenario and also the so called enhanced neutrino 
emission which
results in a much lower temperatures in young NSs.
In order to avoid dealing with the uncertainties related to the behavior
of the field within the superconducting core, as a first step in the present 
paper, 
we consider
only magnetic fields not penetrating the NS core.
Thus, the present analysis is within the framework of the 
crustal magnetic field hypothesis.
We use our results from the GR treatment 
of thermal and magnetic field evolution implemented
by a semi-relativistic treatment of the spin evolution 
to confront the crustal field hypothesis 
to observational data.

The paper is organized as follows: 
In the subsequent section we remind the reader of the 
GR formulation of the equations of stellar 
structure, the heat transport and conservation equations
as well as the induction equation
on a static spherically symmetric background geometry.
We also present  the evolution equations for an axisymmetric dipolar 
magnetic field as well as relativistic expressions for the Joule heating. 
Section 3 describes the microphysics  used in our models.
In section 4 we present the results of our model calculations and 
section 5 is devoted to the discussion 
of the results.

\section{General Relativistic formalism of static spherical stars}

The spacetime geometry will be assumed to be spherically
symmetric, which
means that our results may not be accurate for fast rotating
neutron stars.
Employing the familiar Schwarzschild coordinates $(t,r,\theta,\phi)$,
the interior and exterior spacetime geometry  takes the form
(Misner et al. 1973\cite{MTW73}; Wald 1984\cite{W84})
\be
ds^2 = - e^{2 \Phi} \; c^2 dt^2 + \frac{dr^2}{1 - 2 G m/c^2r} +
 r^2 d \Omega ^2.
\label{equ:schwarzschild}
\ee
The radial proper length is thus $ dl = dr/\sqrt{1 - 2 G m/c^2r}$
and the proper time $d{\tau} = e^{\Phi} dt$.

Einstein's equations coupled to a perfect fluid energy-momentum tensor
give us the standard equations (Misner et al. 1973\cite{MTW73}; 
Wald 1984\cite{W84}):
\be
\frac{dm}{dr} = 4 \pi r^2 \rho,
\label{equ:emass}
\ee
which determines the so called mass function $m=m(r)$,
\be
\frac{d\Phi}{dr} = \frac{Gmc^2 + 4 {\pi} G r^3 P}{c^4r^2(1-2 G m/c^2r)},
\label{equ:phi}
\ee
for the `gravitational potential' $\Phi = \Phi(r)$
and the TOV equation of hydrostatic equilibrium
\begin{displaymath}
\frac{dP}{dr} = -({\rho}c^2 + P)\frac{d\Phi}{dr} =
\end{displaymath}
\be
     -\frac{(\rho + P/c^2)(Gm + 4 \pi G r^3 P/c^2)}{r^2(1-2Gm/c^2r)}.
\label{equ:TOV}
\ee
Regularity of the geometry at $r=0$ implies the inner boundary condition
for Eq.~\ref{equ:emass}
\be 
m(r=0) = 0
\ee
while for Eq.~\ref{equ:TOV} the central pressure
\be
P(r=0) = P_c
\ee
is specified, through the EOS of the form $P=P(\rho)$,
with the central density $\rho_c$ treated as a free parameter.
Due to the linear nature of Eq.(3),  $\Phi$ can be scaled so that it always
can be arranged to fulfill:
\be
e^{\Phi(R)} = \sqrt{1 - \frac{2GM}{c^2R} },
\ee
Once the interior spacetime geometry has been so specified it 
is joined smoothly across the "surface" of the star
to an exterior Schwarzschild field  characterized by $M=m(R)$.

It should be mentioned that the stellar surface in our computation
is fixed by
\be
R = R_{\rm star} = r(\rho = \rho_b)
\ee
where $\rho_b = 10^{10}$ g cm$^{-3}$.
This guarantees that the EOS is temperature independent.
The layers at densities below $\rho_b$, called the {\em envelope},
are treated separately (see \S~\ref{sec:bound}).

\subsection{Thermal evolution equations}

Besides the above equations of stellar structure we shall need 
the equations describing the thermal evolution of the star.
At the temperatures we are interested in, the neutrinos have a mean free path 
much 
larger than the radius of the star (Shapiro \& Teukolsky 1983\cite{ST83}) and 
thus 
leave the star once they are produced.
Energy balance arguments (see for instance Thorne (1966\cite{T66}) then imply
\be
\frac{d(L e^{2 \Phi})}{dr} = -\frac{4 \pi r^2 n e^{\Phi}}{\sqrt{1 - 2Gm/c^2r}}
      \, \left( \frac{d\epsilon}{dt} + e^{\Phi} (q_{\nu}-q_h) \right) 
\label{Equ2-5}
\ee
where $L$ is the internal luminosity,
$\epsilon$ the internal energy per baryon,
$q_{\nu}, q_h$ is the neutrino emissivity and  heating rate {\em per baryon}
while $n$ stands for the baryon number density.
The corresponding inner boundary condition for $L$ is
\be
L(r=0) = 0
\ee
The time derivative of $\epsilon$
can be written in the form
\be
\frac{d \epsilon}{dt} = \frac{d \epsilon}{dT} \cdot \frac{dT}{dt}
                      =         c_v           \cdot \frac{dT}{dt} \;
\ee
through the specific heat at constant volume $c_v$
(which, for degenerate matter, is the same as the specific heat 
at constant pressure $c_P$).

The energy transport equation is~:
\be
\frac{d(T e^{\Phi})}{dr} = - \frac{3}{16 \sigma_{SB}} \frac{\kappa \rho}{T^3}
   \frac{L e^{\Phi}}{4 \pi r^2 \sqrt{1 - 2Gm/c^2r}}
\label{Equ2-6}
\ee
where $T$ is the local temperature, $\sigma_{SB}$ the Stefan-Boltzmann
constant and $\kappa$ the `opacity'.
(Notice that within the relativistic framework
an `isothermal' configuration is defined by
$e^{\phi} \cdot T = {\rm constant}$ instead of $T$ = constant.)
The associated boundary condition is
\be
T_b = T_b(L_b)
\label{equ:tb-te}
\ee
which relates the temperature at the outer boundary (defined more precisely 
further bellow,), $T_b$, to the
luminosity, $L_b$, in this layer.
The location of this outer boundary layer is chosen such that $L_b$ be equal 
to the total photon luminosity of the star, $L_* \equiv L(r=R)$, 
which in turn is related to the effective temperature
$T_e$ by $L_* \equiv 4 \pi R^2 \sigma_{SB} T_e^4$.
We can thus write Eq.~\ref{equ:tb-te} as $T_b = T_b(T_e)$ and
this `$T_b$ - $T_e$ relationship' is discussed in Sect.~\ref{sec:bound}.

The opacity is related to the total (thermal) conductivity by:
\be
\lambda = \frac{16 \sigma_{SB} \; T^3}{3 \kappa \rho}.
\label{Equlambdakappa}
\ee
If one neglects the red-shift $e^{\Phi}$ and defines the energy flux $F$ as
$L/(4 \pi r^2)$ one can write Eq.~\ref{Equ2-6} as
\be
F = - \lambda \cdot \nabla T
\label{equ:HeatCond}
\ee
where $\nabla$ is the radial gradient calculated with the proper length, which
is the usual form of the heat conduction equation. The total conductivity is
the sum of the electron and photon conductivities
\be
\lambda = \lambda_e + \lambda_{\gamma}
\ee
since these two processes of heat conduction work independently and in
parallel.

We will present our results of thermal evolution by using the
`effective temperature at infinity' 
$T_e^{\infty} \equiv T_e \cdot e^{\Phi(R)}$
related to the `luminosity at infinity'
$L_*^{\infty} \equiv L_* \cdot e^{2 \Phi(R)}$
through the `radius at infinity'
$R^{\infty} \equiv R \cdot e^{-\Phi(R)}$
by 
\be
L_*^{\infty} = 4 \pi (R^{\infty})^2 \sigma_{SB} (T_e^{\infty})^4.
\ee
These three quantities `at infinity' are, in principle, measurable
and, in particular, $R^{\infty}$ would be the areal radius of the
star that an observer `at infinity' would measure with an extremely
high angular resolution instrument (Page 1995\cite{P95}).

The solution of the complete set of equations of stellar structure,
Eqs.~(2-7) and
thermal evolution, Eqs.~(9-13), requires knowledge of the 
equation of state, the opacity $\kappa$, the
neutrino emissivity $q_{\nu}$ and also the specific heat $c_v$.
We shall devote Sect.~\ref{sec:phys} to the detailed specification of those 
variables.

\subsection{Magnetic evolution equations}

Besides the equations of stellar structure and thermal evolution we
shall also need the equations describing the magnetic field evolution.
In this paper consideration will be restricted to dipolar (poloidal) magnetic 
fields.
The GR formulation of the evolution equation of such fields
has been discussed in detail by GPZ00\cite{GPZ00} while for a
more general set up see R\"adler et al. (2000\cite{RFGZ00}).
As has been shown in the first  reference, such a field can be expressed in 
terms of a 
relativistic generalization of the familiar  Stoke's stream function by: 
\be
B^{r}(t,r,{\theta})=\frac{2\, F(t,r)}{r^2}\; \cos{\theta},
\ee
\be
B^{\theta}(t,r,{\theta})=
- {1\over r}
\left(1-{2Gm\over c^2r}\right)^{1\over 2}
{\partial F(t,r) \over {\partial r}}\; \sin{\theta}.
\label{equ:stoke}
\ee
while  the relevant induction equation (see Appendix) yields the following 
equation
for the relativistic Stoke's function 
%
\begin{displaymath}
{4{\pi}{\sigma}\over c^{2}}e^{-{\Phi}}{\partial {F}\over {\partial t}}=
\left(1-{2Gm\over c^{2}r}\right){\partial^{2}F \over {\partial r^{2}}}
\end{displaymath}
\be
+{1\over r^{2}}{\partial F\over {\partial r}}
\left[
{{2Gm\over c^{2}}+{4{\pi}G\over c^{2}}r^{3}}\left({P\over c^{2}}-{\rho}\right)
\right]
-{2\over r^{2}}F.
\label{equ:IndequF}
\ee
The appropriate boundary conditions, as $r \rightarrow 0$,
is the same as in the flat space case: a regular  field at the star's center 
requires 
${F(t,r)\over r^{2}}$ to be finite.
The outer boundary condition, however, differs from that valid in the 
flat space, and its GR form is as follows
(see: GPZ00\cite{GPZ00}):
\be
\left.R{\partial {F(t,r)}\over {\partial r}}\right|_{R}=G(y)
{F(t,R)}
\label{equ:outbc_a}
\ee
where:
\be
G(y)=y{{2y \ln(1-y^{-1})+{2y-1\over y-1}}\over y^{2} \ln(1-y^{-1})+
y+{1\over 2}}   \;\;\;\; {\rm with} \;\;\;\; y=R/R_S      
\label{equ:outbc_b}
\ee
($R_S \equiv 2GM/c^2$ being the star's Schwarzschild radius).
Since $G(y) < 0$ (in particular, in the flat space-time case, $G(\infty) = -1$)
the boundary condition forces a bending of $F$ in the upper layers.

As an initial profile for the Stoke function we use the same formula as UK97,
for later comparison with the work of these authors:
\be
F(r,0)=B_0 R^2 \; (1-r^2/r_0^2)/(1-R^2/r_0^2) \;\;\;\; {\rm at} \;\;\; r>r_0
\label{equ:initial_F}
\ee
\begin{displaymath}
F(r,0)=0 \;\;\;\;\;\;\;\;\;\;\;\;\;\;\;\;\;\;\;\;\;\;\;\;\;\;\;\;\;\;\;\;\;\;
           \;\;\;\;\;\;\;\;\;\;\;\;\;\;\;\; {\rm at} \;\;\; r<r_0
\end{displaymath}
Notice that this initial $F(r,0)$ does not satisfy the outer boundary 
condition,
but will immediately be forced to do it at the first numerical time step.
There will thus be a rapid relaxation of $F$ in its early evolution due to
the enforcement of the boundary condition and the propagation of the
resulting curvature of $F$ toward higher densities.

\subsection{Joule heating}

Due to the finite conductivity $\sigma$, magnetic energy is 
dissipated into heat 
(= Joule heating).
The heat production per unit (proper) time and 
unit (proper)
volume is given by
\be
Q_h = n \cdot q_h= \frac{\vec j^2}{\sigma}
\label{equ:joule}
\ee
where $\vec j$ is the current.
As it is shown in the Appendix, for a dipole poloidal magnetic field $\vec B$, 
$\vec j$ is given by $\vec j = j_{\phi} \vec e_{\phi}$ with
\begin{displaymath}
j_{\phi} =
\frac{c}{4 \pi} \frac{\sin{\theta}}{r} \; \times
\end{displaymath}
\be
\left\{\! e^{-\Phi} \! \!
\left(1-\frac{2Gm}{c^2r}\right)^{\! \frac{1}{2}} \! \! 
\frac{\partial}{\partial r} \! \!
\left[e^{\Phi} \! \left(1-\frac{2Gm}{c^2r}\right)^{\! \frac{1}{2}} 
     \! \!  \frac{\partial F}{\partial r}\right] - 
   \frac{2F}{r^2} \! \right\}.
\label{equ:jouleheatrel}
\ee
It is seen immediately 
that in the limit $e^{\Phi}=1$ and $m=0$ the above expression 
reduces to its  flat
space-time counterpart (Miralles et al. 1998\cite{MUK98}).
Since our numerical calculations assume spherical symmetry we
use, in Eq.~\ref{equ:jouleheatrel}, a spherical average of
$\sin^2 \theta$ ($<\sin^2 \theta> \; = \; \frac{2}{3}$) obtaining
\begin{displaymath}
<Q_h> = \frac{c^2}{24 \pi^2}\frac{1}{\sigma}\frac{1}{r^2} \times
\end{displaymath}
\be
\left\{\! e^{-\Phi} \! \!
\left(1-\frac{2Gm}{c^2r}\right)^{\! \frac{1}{2}} \! \! 
\frac{\partial}{\partial r} \! \!
\left[e^{\Phi} \! \left(1-\frac{2Gm}{c^2r}\right)^{\! \frac{1}{2}} 
     \! \!  \frac{\partial F}{\partial r}\right] - 
   \frac{2F}{r^2} \! \right\}^2.
\label{equ:jouleheatrel_av}
\ee

\subsection{Outer boundary: magnetized envelopes
            \label{sec:bound}}
The layers at densities below $\rho_b = 10^{10}$ g cm$^{-3}$ are defined as 
the {\em envelope},
and extend up to the {\em atmosphere}, where the {\em photosphere} is located,
while by {\em interior} we mean the whole star where $\rho > \rho_b$.
The presence of the magnetic field affects strongly the heat transport in
the envelope (but not in the deeper layer where $\rho > 10^{10}$ g cm$^{-3}$,
which motivates the above choice of $\rho_b$), and
results in a non uniform distribution of the surface temperature.
The corresponding `$T_b - T_e$ relationships' have been calculated by 
Page \& Sarmiento (1996\cite{PS96}) for dipolar (and quadrupolar) fields.
We use these results which hence give us a field dependent `$T_b - T_e$
relationship' that adjust itself to the evolution of the magnetic field.
Notice however that, for fields much stronger than 10$^{12}$ G, this 
relationship is not reliable when $T_e$ is much lower than 10$^6$ K,
and it is most probably very inaccurate when $T_e$ is below 10$^5$ K.

Our outer boundary condition assumes that the
envelope is made of 
catalyzed matter. 
If an upper layer of light elements were present, the heat transport is
strongly enhanced when no magnetic field is present 
(Potekhin, Chabrier, \& Yakovlev 1997\cite{PCY97}), but there is, to date,
no published model of magnetized envelopes with light elements.

One should finally emphasize that
when the thermal evolution is controlled by the Joule heating
the star's luminosity, and $T_e$, is given by the heating rate and is 
independent of the outer boundary condition as discussed at the end
of \S~\ref{sec:evol-stand}.

This closes the system of equations and boundary conditions to be solved.

\subsection{Numerical method}

The thermal evolution equations are solved by a Henyey-type code 
(e.g., Page 1989\cite{P89}) while the induction equation for the
Stoke function is solved with a Crank-Nicholson method
(Press et al., 1986\cite{PFTV86}).
The whole set of equations for the thermo-magnetic evolution should be
solved simultaneously at each time step but we have decided to solve the
thermal and then the magnetic equations alternatively, i.e., the thermal
equations are solved at a given time step using the field of the previous
step (which appears in the Joule heating term) and once the new temperature
is obtained the induction equation is solved to obtain the new magnetic field.
This method is much faster than a full simultaneous solution and gives results 
which, as we have verified explicitly in a few case, are practically
indistinguishable from the full simultaneous solution.

\section{Input microphysics \label{sec:phys}}

\subsection{The equation of state}

The first ingredient needed to build NS models is the equation of state (EOS).
In principle the EOS should give us not only the relationship between pressure
and density, i.e., $P = P(\rho)$, but also the chemical composition of matter.

We separate the crust from the core at the density 
$\rho = \rho_{cr} \equiv 1.6 \times 10^{14}$ g cm$^{-3}$
(Lorenz, Ravenhall \& Pethick 1993\cite{LRP93})
and use the EOSs of Negele \& Vautherin (1973\cite{NV73}) for the inner crust,
at $\rho > \rho_{drip} \equiv 4.4 \times 10^{11}$ g cm$^{-3}$,
and Haensel, Zdunik \& Dobaczewski (1989\cite{HZD89}) for the outer crust,
i.e., we assume that the chemical composition is that of cold catalyzed matter.
The EOS, and its associated chemical composition, in the crust is well
determined under the assumption that matter is in its (catalyzed) ground state.
There is however still the possibility that a strong phase of hypercritical
accretion occurred after the supernova explosion, which may, or may not, alter
 the 
chemical composition of the crust.
We will not consider this possibility here but only mention that it does have
an enormous effect on the magnetic field and its subsequent evolution
(Geppert, Page \& Zannias 1999\cite{GPZ99})

In the core, the EOS is relatively well constrained up to densities
around 2 - 3 $\times \rho_{cr}$ while its behavior at higher densities
is still a mystery (Prakash 1998\cite{Pr98}).
We will thus consider three different cases which hopefully illustrate the 
whole range of possibilities.
We take as a `Medium EOS' the one calculated by Wiringa, Fiks \& Fabrocini 
(1988\cite{WFF88}), using their model called av14+UVII, 
a `Stiff EOS' from Pandharipande, Pines \& Smith (1976\cite{PPS76})
and a `Soft EOS' from Pandharipande (1971\cite{P71}).
Notice that the Stiff EOS is based on the presence of a lattice of neutrons
in the inner core and is not anymore considered as realistic but we still
use it since it has been used by many authors and represents the case of
extreme stiffness.
Moreover, none of the consequences of the presence of such a lattice phase
is taken into account in our calculations, e.g., in $C_v$, $\epsilon_{\nu}$
and the transport coefficients: we will assume that neutrons and protons 
form a quantum liquid and also boldly consider them as superfluid and
superconductor. 
Our opinion is that this EOS should be abandoned but we consider it
for comparison with previous works of other authors.
The Soft EOS on the other hand, despite of its age, is still representative
of modern soft EOSs and includes hyperons.

We will consider models of 1.4 \Msun stars whose overall properties are 
listed in Table~\ref{ta:stars}.

\begin{center} 
\begin{table}[t] 
\begin{tabular}{cccccc}
EOS    & Radius &   Crust   &  Crust   &      Central    &      Moment  \\
       &        & thickness &  mass    &      density    &     of  inertia \\
       &  [km]  &   [km]    & [\Msun]  & [10$^{15}$ &  [10$^{45}$  \\
       &        &           &          &    g/cm$^3$]  & g cm$^2$]  \\
\hline
Stiff  & 14.94  &   1.52    &  0.069   &        0.49     &       1.83     \\
Medium & 10.48  &   0.72    &  0.022   &        1.17     &       1.08     \\
Soft   &  7.1   &   0.22    &  0.003   &        5.59     &       0.72     \\
\hline
\end{tabular}
\caption[Stars' properties]
         {Properties of the 1.4 \Msun neutron star models
         \label{ta:stars}}
\end{table}
\end{center}

\vspace{-1cm}
\subsection{Neutrino processes}

Neutrino emission drives the cooling as long as the internal temperature
is higher than about $10^8$ K.
For processes in the crust we consider the two dominant ones which are
the plasmon process  and the electron-ion bremsstrahlung (Page 1989\cite{P89}).
The former is only relevant during the first few years of the life of the 
NS but is very strong and brings down the crust temperature to about 
$10^9$ K, relaxing it from the arbitrary initial conditions.
The latter has only a very small effect, mostly when the surface temperature
is around $10^6$ K, i.e., for young stars.

The crucial neutrino emission processes occur in the core and we consider
two scenarios (see Page 1998\cite{P98} for more details).
In the `Standard Cooling', or `slow cooling', scenario we include the modified
Urca processes and their associated, and weaker, bremsstrahlung processes
following Yakovlev \& Levenfish (1995\cite{YL95}).
This scenario applies when the NS core contains only neutrons, protons, as 
well as
electrons and muons which maintain charge neutrality, and the proton
fraction is low enough that the direct Urca process is forbidden by momentum
conservation.
In the `Enhanced Cooling', or `fast cooling', cases we add a strong neutrino 
emission at densities larger to $\rho_{fast} = 4 \times 10^{14}$ g cm$^{-3}$
with a rate
\be
\epsilon_{\nu}^{FAST} = 
10^{26} \; (\rho/\rho_{cr})^{2/3} \; T_9^6 \;\; {\rm erg \; g^{-1} \;  s^{-1}}
\ee
where $T_9 \equiv T/10^9$ K.
This rate is representative of many of the possible enhanced ones as the 
direct Urca from nucleons or hyperons, but is stronger than what produced
by a pion or kaon condensate (Prakash 1998\cite{Pr98}).
For, comparison the inefficient modified Urca process gives approximately
$\epsilon_{\nu}^{MURCA} \sim 
10^{21} \; T_9^8 \;\; {\rm erg \; g^{-1} \;  s^{-1}}$.

In our three model 1.4 \Msun stars, the masses of the inner cores where the
fast neutrino emission is allowed for the `Enhanced Cooling' cases are,
0.11, 1.29 and 1.39 \Msun for the Soft, Medium and Stiff EOS, respectively.

\subsection{Electrical conductivities}

\begin{figure}
\begin{center}
\resizebox{\hsize}{!}{\includegraphics{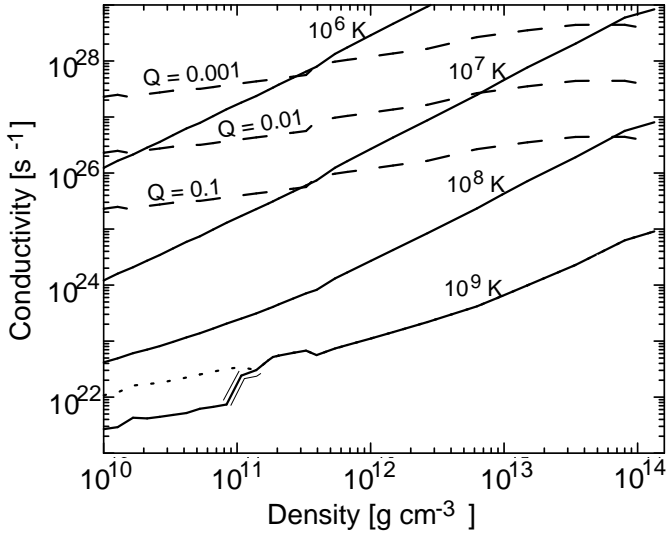}}
\caption{Electrical conductivities in the crust.
         The dashed lines show $\sigma_{\rm imp}$, which is
	 temperature independent, for three different values
         of the impurity content $Q_{\rm imp}$.
	 The continuous lines show $\sigma$ for electron-phonon scattering,
         in the solid phase, and electron-ion scattering, in the liquid phase,
	 at four different temperatures.
	 Notice that these four curves all correspond to the solid
	 phase except for the $10^9$ K one where the low density part
	 is still in the liquid phase: the dotted extension show the value
	 $\sigma$ would have if the matter were in the solid phase and the
	 short tripped segment indicates the region where $\sigma$ is 
         interpolated
	 between the electron-phonon and the electron-ion values.
	 Finally, the increase of $\sigma$ in this $10^9$ K case at densities
	 just above the transition density, i.e., temperatures just below
	 the melting temperature, is due to the Debye-Waller factor
	 (Itoh et al 1984\protect\cite{IKMS84}).
	 For comparison with the cooling curves, notice that interior 
         temperatures 
	 of $10^6$, $10^7$, $10^8$, and $10^9$ K 
	 roughly correspond to effective temperatures around
         $10^5$, $3 \times 10^5$, $10^6$, and $3 \times 10^6$, respectively.}
\label{fig:sigma}
\end{center}
\end{figure}

In general, the electrical conductivity $\sigma$ can be expressed in terms of 
the electron relaxation time $\tau$ as
\be
\sigma = \frac{e^2 n_e \tau}{m_e^*},
\label{equ:sigma}
\ee
where 
\be
m_e^* = \mu_e / c^2 = m_e \, [1 + 1.018 \, (\rho_6 \, Z/A)^{2/3}]^{1/2}
\label{equ:mue}
\ee
is the electron effective mass and $n_e$ the electron number density
($\mu_e$§ being the electron chemical potential, 
$\rho_6 = \rho / 10^6$ g cm$^{-3}$,
Z and A the charge and mass number of the ions).
In the liquid phase we use the calculation of $\tau = \tau_{\rm e-i}$ for 
electron-ion scattering by Itoh et al (1983\cite{IMII83}).
In the solid phase $\tau$ is given by
\be
\frac{1}{\tau} = \frac{1}{\tau_{\rm e-ph}} + \frac{1}{\tau_{\rm e-imp}}
\label{equ:tau}
\ee
where we use the results of Itoh et al (1984\cite{IKMS84}) for the 
electron-phonon scattering
$\tau_{\rm e-ph}$ and of Yakovlev \& Urpin (1980\cite{YU80})
for the electron-impurity scattering $\tau_{\rm e-imp}$.
In general, $\tau_{{\rm e-i}}$ is a function of $\rho$, A and Z, and
$\tau_{\rm e-ph}$ depends also on the temperature $T$ while
$\tau_{\rm e-imp}$ depends on $\rho$, $Z$ and the impurity concentration 
$Q_{\rm imp}$.
We show in Fig.~\ref{fig:sigma} the value of $\sigma$ for typical values of 
$T$ and $Q_{\rm imp}$.

\subsection{Pairing}

\begin{figure}
\begin{center}
\resizebox{4in}{!}{\includegraphics{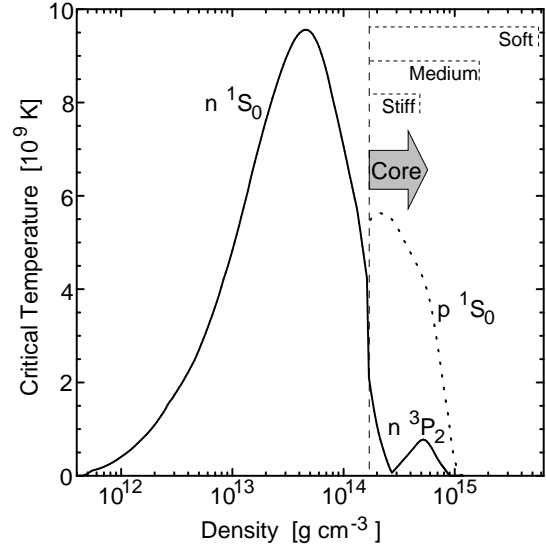}}
%
\caption{Pairing critical temperatures adopted in this work, for 
         neutrons in the $^1$S$_0$ state 
         (from Ainsworth, Wambach \& Pine 1989\protect\cite{AWP89})
         and $^3$P$_2$ state
         (from Takatsuka 1972\protect\cite{T72})
         and protons in the $^1$S$_0$ state 
         (from Baldo et al. 1992\protect\cite{BCLL92}).
	 The dashed lines show the central densities of our three model 
	 1.4 \Msun stars with the soft, medium and stiff EOS.}
\label{fig:Tc}
\end{center}
\end{figure}

The occurrence of pairing, neutron superfluidity and proton superconductivity, 
strongly affects both the thermal and magnetic evolution of neutron stars.
The thermal effects are very strong during the neutrino cooling phase,
which last about 10$^5$ to almost 10$^7$ yrs depending on the model, and 
the subsequent photon cooling phase (see, e.g., Page 1998\cite{P98} for a 
review).
As a result, pairing will in a large part determine the time at which 
Joule heating starts to control the thermal evolution and then during this 
Joule heating phase the effect of pairing becomes negligible.
We treat the suppressive effect of pairing on $C_v$ and $\epsilon_{\nu}$
according to the treatment of Levenfish \& Yakovlev (1994a\cite{LY94a},
1994b\cite{LY94b}) and Yakovlev \& Levenfish (1995\cite{YL95}).

The superconductive phase in the core has the effect of producing an 
impenetrable
barrier for the magnetic field, which is initially confined to the crust in
the models of the present work.
This guarantees the confinement of the magnetic field, and the currents,
to the crust since the superconductive phase transition happens, at
the crust-core boundary, well before the magnetic field had time to diffuse
to this layer.
For definiteness we plot in Fig.~\ref{fig:Tc} the pairing critical temperatures
that we adopt in this paper.
Notice that the values we adopt for $^1$S$_0$ pairing of both neutrons and 
protons
are typical of modern calculations.
In the case of neutron $^3$P$_2$ pairing we explicitly adopt values which 
vanish
a densities above 10$^{15}$ g cm$^{-3}$ to ensure that pairing will not 
suppress
the strong neutrino emission in our Enhanced Cooling scenarios for the
Soft and Medium EOSs.

In the case of the Stiff EOS any published calculation of neutron or proton 
pairing 
shows a non vanishing value of $T_c$ at the density in the center of the star,
 given
the low value of this central density.
To avoid suppression of the neutrino emission by pairing we will assume that 
{\em the fast neutrino emission is not affected by neutron and proton pairing
in this case of Stiff EOS in the Enhanced Cooling scenario }.

\begin{center} 
\begin{table}[t] 
\begin{center} 
\begin{tabular}{ccccc}
Model   &    $B_0$ [G]    &   $\rho_0$ [gm/cm$^3$]      &  $Q_{\rm imp}$  \\
        &                 &                             &                 \\
\hline
        &                 &                             &                 \\
   1    &  $10^{11}$      &        $10^{12}$            &  0.1            \\
   2    &  $10^{12}$      &        $10^{13}$            & 0.01            \\
   3    &  $10^{13}$      &        $10^{14}$            & 0.001           \\
        &                 &                             &                 \\
\hline
\end{tabular}
\caption[Initial Field]
         {Initial magnetic field configurations and impurity concentrations in
          our three classes of models.
         \label{ta:field}}
\end{center}
\end{table}
\end{center}

\section{Results}

We will present here our results for the thermal, magnetic and rotational
evolution of isolated neutron stars considering 1.4 \Msun stars built with
the three EOSs described above within both the `standard' and the `fast'
cooling scenarios.
In order to investigate the influence of the (a priori unknown) initial 
structure and 
strength of the magnetic field onto NS's evolution, we considered for each EOS
and cooling scenario three classes of qualitatively different field models, 
existent at the beginning of the evolution. 
They are characterized by the initial surface field strength $B_0$, depth of 
penetration
of the current $\rho_0$ and impurity concentration $Q$ as listed in 
Table~\ref{ta:field}.

\begin{figure}
\begin{center}
\resizebox{4in}{!}{\includegraphics{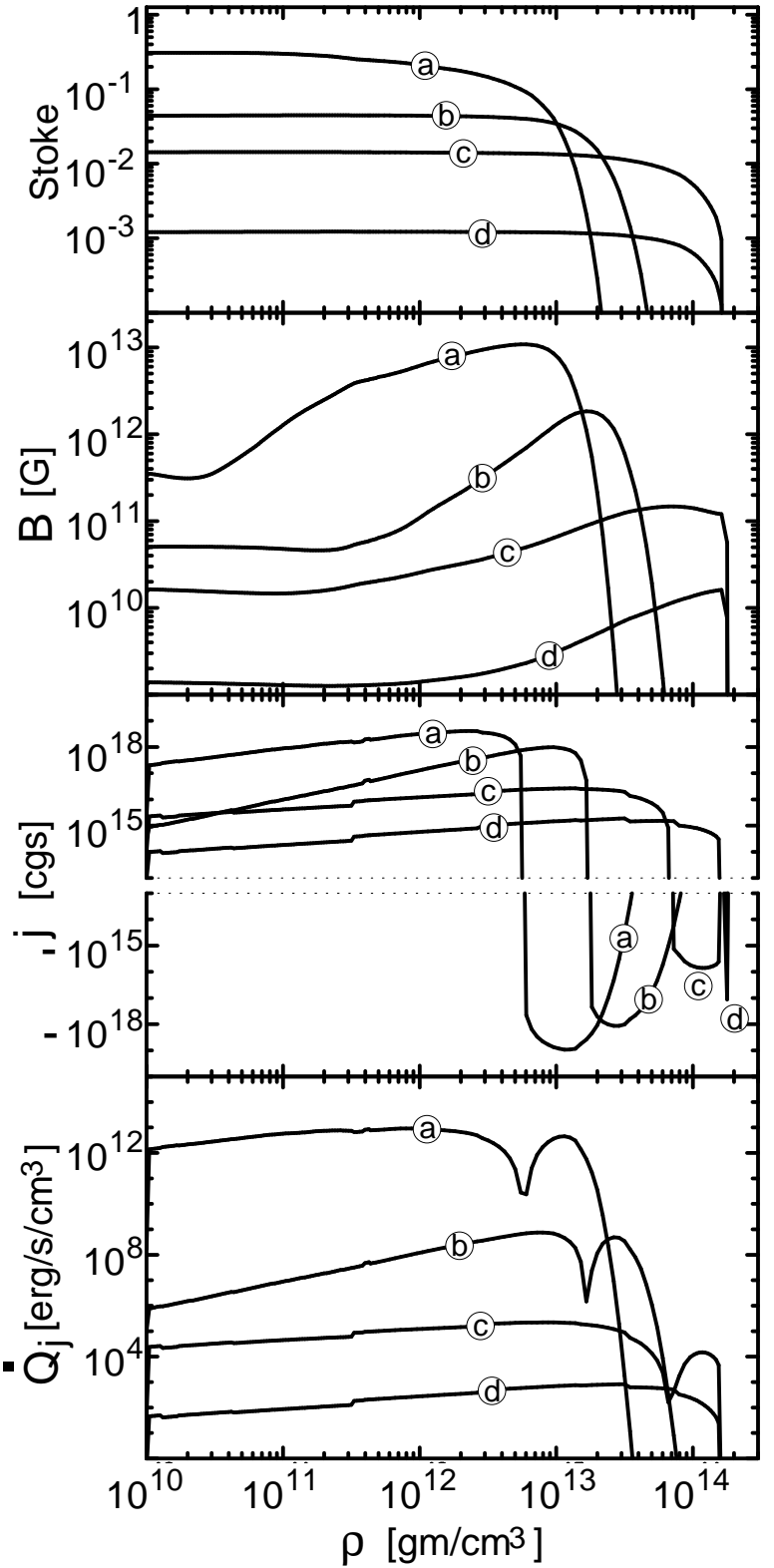}}
\caption{Stoke function, magnetic field, $B$, currents, $j$ and 
         Joule heating rate, $\dot{Q}_j$ in the crust, for model 2 of the 
         central panel of Fig.~\protect~\ref{fig:cool_field_slow}.
         See text, \S4.1, for details.}
\label{fig:sbbch}
\end{center}
\end{figure}

\subsection{A detailed example}

\begin{figure}
\begin{center}
\resizebox{2.5in}{!}{\includegraphics{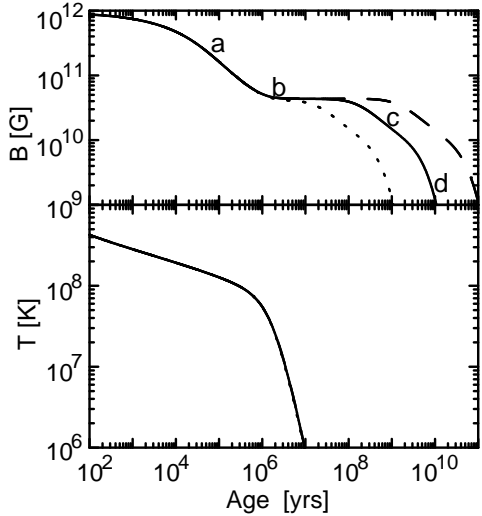}}
\caption{Evolution of the internal temperature (lower panel) and surface
        magnetic field for the model described in Fig.~\protect\ref{fig:sbbch},
	 continuous line. The two models in dashed and dotted lines show the
	 corresponding evolution for impurity contents $Q_{\rm imp} =$ 0.001
	 and 0.1, respectively (the continuous line has $Q_{\rm imp} =$ 0.01). 
         Notice that in these cases the heating is so small
	 that the three models follow the same thermal evolution track.}
\label{fig:bt-det}
\end{center}
\end{figure}

\begin{figure*}
\begin{center}
\resizebox{\hsize}{!}{\includegraphics{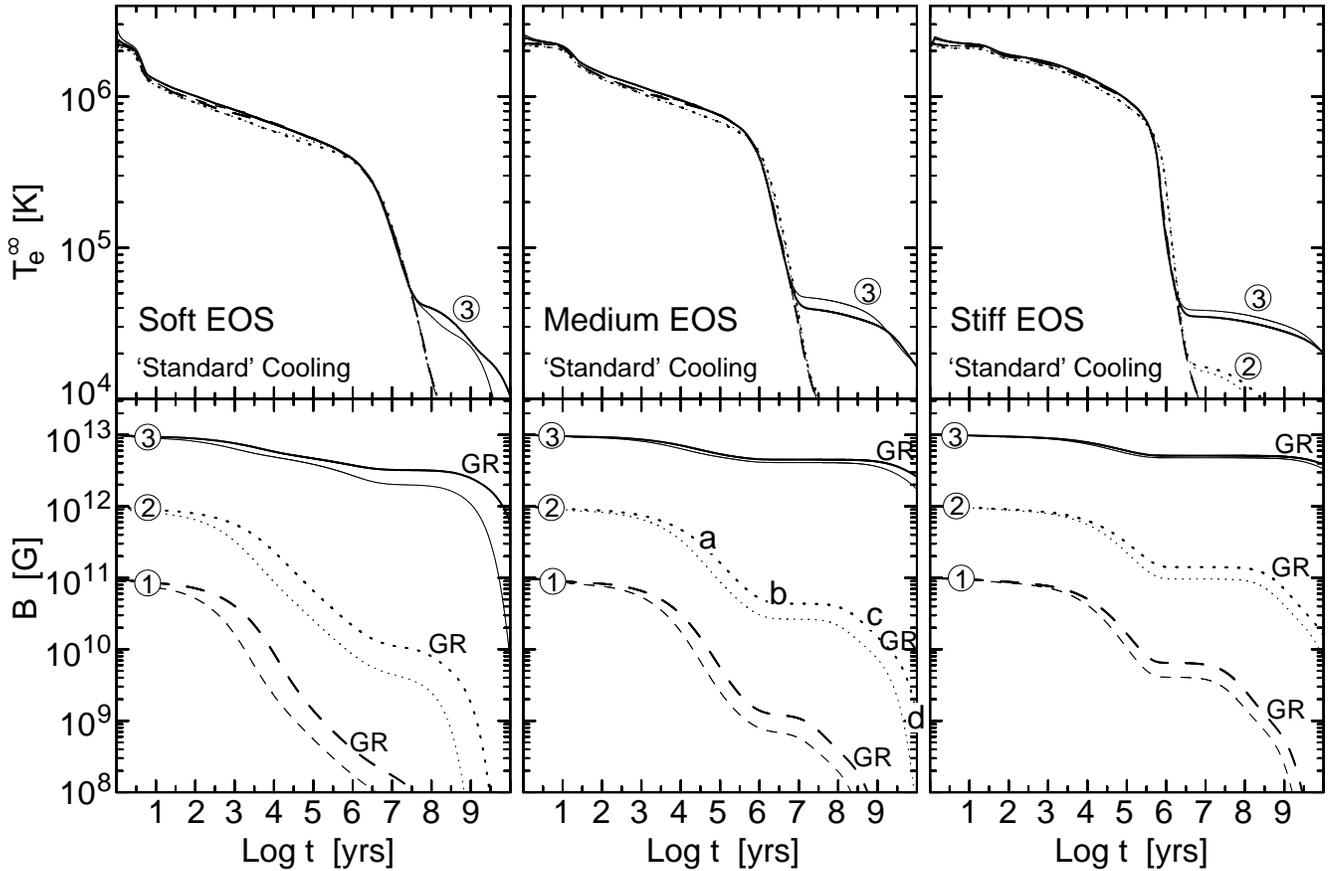}}
\caption{Thermal and magnetic evolution within the `standard' cooling scenario.
         Evolutionary curves are labeled following notations of
	 table~\protect\ref{ta:field}. 
         See text for details.
        }
\label{fig:cool_field_slow}
\end{center}
\end{figure*}

We first discuss here an example of the internal evolution of the Stoke 
function,
magnetic field, currents and local heating rate, as shown in 
Fig.~\ref{fig:sbbch},
which will help for the general discussion presented below.
We choose the 1.4 \Msun star with the medium EOS and `standard' cooling;
the initial surface field strength is 10$^{12}$ G and the currents are
initially located at $\rho_0 = 10^{13}$ g cm$^{-3}$.
The impurity content for $\sigma_{\rm imp}$ is $Q_{\rm imp} = 0.01$.
This is the model 2 of the central panel of Fig.~\ref{fig:cool_field_slow}
but it is reproduced in Fig.~\ref{fig:bt-det} along with two similar
models with different impurity contents (upper panel) and the evolution
of the internal temperature (lower panel).
GR effects are included but will be discussed in the next subsections.

The four chosen times, labeled as a, b, c \& d are marked in the upper panel
of Fig.~\ref{fig:bt-det} and correspond, respectively, to
a) the initial field decay during the period when the neutron star is still hot
   and $\sigma$ is temperature dependent, being dominated by electron-phonon
   scattering,
b) the plateau where the cooling lead to an enormous increase of $\sigma$,
   and consequently a stagnation of the field decay, which will eventually
   be controlled by impurity scattering,
c) second phase of decay of the field when time becomes comparable to the 
impurity scattering
   decay time scale and, finally,
d) the late exponential decay.

The four panels of Fig.~\ref{fig:sbbch} directly illustrate the diffusion of 
the field
toward regions of higher conductivity as time runs.
Once $\sigma$ becomes $T$ independent the diffusion equation formulates an
eigenvalue problem whose solution can be formally written as
$F(\rho,t) = \sum_{n=1}^{\infty} \exp(-t/\tau_n) a_n F_n(\rho)$, in terms of 
eigenmodes
$F_n$ with decay times $\tau_n$, and expansion coefficients $a_n$.
Since the n-th mode, $F_n(\rho)$, has n nodes, in the crust, initially a very 
large number
of modes must contributes significantly, i.e., with large positive and 
negative 
coefficients $a_n$, to produce a mutual cancellation resulting in a vanishing 
$F(\rho,t)$ in the high density region.
The diffusion of the stoke function into the high density region is simply due
to the faster decay of the modes with $n > 1$ compared to the nodeless 
fundamental mode.
When the fundamental mode is dominating, i.e., when the stoke function is non 
zero
in the whole crust (it has reached the crust-core boundary through diffusion),
the field evolution is a power-law like decay, phase c 
(Urpin, Chanmugan \& Sang 1994\cite{UCS94}). 
Finally, when only the fundamental mode $n = 1$ is left the decay becomes 
purely exponential,
phase d.

Notice that the field strength inside the crust is at least one order of 
magnitude
higher than at the surface: this is due to a very large $B_{\theta}$ forced
by the presence of the derivative $dF/dr$ in Eq.~\ref{equ:stoke}, while at the
surface the boundary condition ensures that $B_r$ and $B_{\theta}$ are 
comparable.

With respect to the currents, noticeable are the negative currents in the 
layers
where the field is growing which are due to induction, i.e., Lenz law.
Once the field has reached the crust-core boundary the currents are positive
in the whole crust and the only negative currents left are the supercurrents
induced in the skin layer of the proton superconductor.
Instead of imposing an ad-hoc boundary condition at the crust-core interface
to simulate the effect of the proton superconductor we have preferred to keep
the central boundary condition and introduce an enormous value for
$\sigma$, 10$^{200}$ s$^{-1}$, once protons become superconductor, i.e.,
when $T < T_c$. 
This allows us to see explicitly the induced supercurrents which, however,
because of the finite radial resolution of the numerical 
scheme, are located in the whole last zone of the crust instead of the
physical skin layer of the superconductor which is a few tens of fermis thick.
We have checked explicitly that an ad-hoc crust-core boundary condition for
$F$ gives the same results as our boundary condition with enormous $\sigma$.

As a last remark, from Fig.~\ref{fig:bt-det}, and comparing with 
Fig.~\ref{fig:sigma},
we see that impurity scattering starts to dominate $\sigma$ after the
field has reached the plateau (phase b):
this stagnation value of the field in independent of $Q_{\rm imp}$ and is
only a result of the enormous increase of $\sigma$ due to the cooling.
After this, the length of the plateau (phase b) is controlled by $Q_{\rm imp}$,
higher impurity contents leading naturally to an earlier onset of the
second phase of field decay (phase c).

\subsection{Magneto-thermal evolution within the `standard' cooling scenario
            \label{sec:evol-stand}}

\begin{figure*}
\begin{center}
\resizebox{\hsize}{!}{\includegraphics{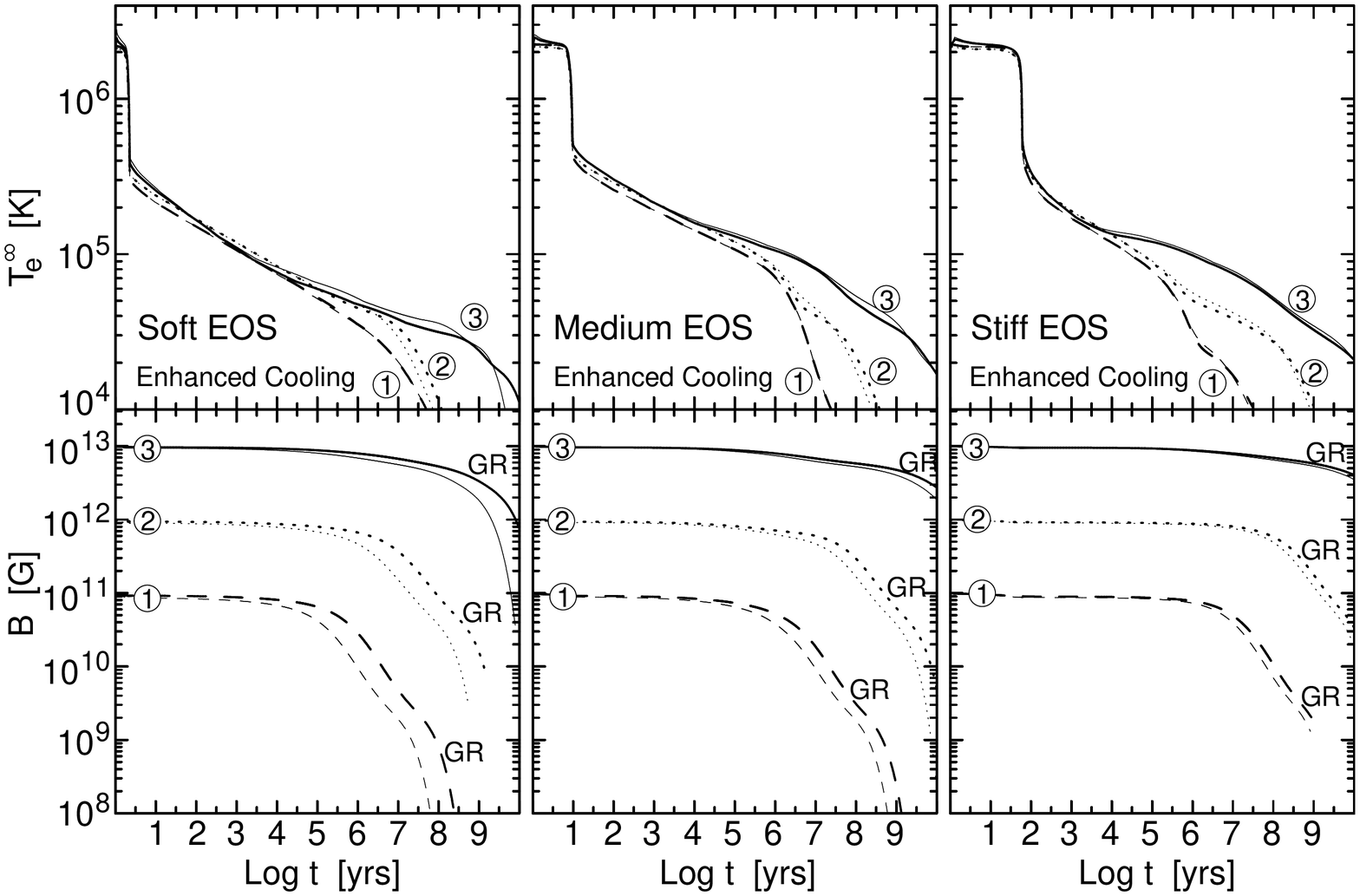}}
\caption{Cooling and field evolution within the `fast' cooling scenario.
         Evolutionary curves are labeled following notations of
	 table~\protect\ref{ta:field}. 
         See text for details.
        }
\label{fig:cool_field_fast}
\end{center}
\end{figure*}

In Figs.~\ref{fig:cool_field_slow} and~\ref{fig:cool_field_fast} we present 
the 
modeled evolution of the, more or less, observable quantities, i.e. the 
surface 
temperature (measured by a far distant observer) and the surface magnetic 
field. 
While the surface temperatures, or, at least, its upper limits, of isolated 
NSs 
can be inferred from X--ray spectra, the magnetic field of isolated pulsars is 
mostly estimated by  the precise measurement of the rotational period and its 
time
derivative.

Within the `standard' cooling scenario the surface temperature stays quite high
($\sim 10^6$ K) during the neutrino dominated cooling era, which lasts 
about $\sim 10^6$ yrs (for our stiff EOS models) to almost $\sim 10^7$ yrs 
(soft EOS models).
Later on the cooling is driven by photon emission from NS's surface which
appears in the cooling curves by a strong increase in the slope.
During the neutrino cooling era, the surface temperature 
drops down by a factor of about $2$ (stiff EOS) to $4$ (soft EOS), i.e. the 
approximately 
isothermal crust has at the end of the neutrino cooling epoch a temperature 
of about 
$5 \cdot 10^7$ (soft EOS) to $10^8$K (stiff EOS). 
As seen from Fig.~\ref{fig:sigma}, for such and higher temperatures, the 
electrical
conductivity in the crust is determined by electron--phonon collisions except 
at the
highest densities.
With the following temperature drop during the photon cooling era,
the electron-phonon relaxation time $\tau_{\rm e-ph}$ increase dramatically 
and thus
impurity scattering will begin to control $\sigma$ (see Eq.~\ref{equ:tau}), 
at a temperature, and thus an age, depending on the impurity concentration.
From that stage on, $\sigma$ becomes temperature independent.

Notice that the differences in the thermal evolution for the different EOSs,
before Joule heating becomes efficient, are mostly due to the differences in 
the
fraction of the core which is paired.
A larger paired region implies a lower neutrino emission, and thus a higher
temperature during the neutrino cooling era, and also a lower specific heat,
and thus an earlier transition to the photon cooling era and a faster 
temperature drop during that era (see, e.g., Page 1998\cite{P98}).
We have taken here the choice of well defined density dependences of $T_c$, for
both the neutrons and protons, independently of the EOS, as shown in 
Fig.~\ref{fig:Tc}.
Different density dependences of the $T_c$'s would obviously give different 
results.
For example, assuming high values of $T_c$ down to center of the star for the 
soft EOS
would results in a cooling history practically indistinguishable form the 
cooling
history of the stiff EOS model.
Given the present uncertainty on the value and density dependence of $T_c$ 
for $^3$P$_2$
neutron pairing (Baldo et al. 1998\cite{BEEHS98}) any choice has, 
unfortunately,
some arbitrariness.
Any effect of the resulting cooling history on the field evolution should 
thus be
specifically formulated in terms of cooling history and not in terms of the
stiffness of the EOS.

We have, very roughly, for the field decay time-scale, 
$\tau_{\rm decay} \propto l^2$, 
$l$ being a typical length scale of the crustal field structure.
This immediately implies that $\tau_{\rm decay}$ increases when $\rho_0$ 
increases 
and also when the stiffness of the EOS is increased, since both increase $l$.
Moreover, since also $\tau_{\rm decay} \propto \sigma$, a higher $\rho_0$ 
locates
the currents in a region of higher $\sigma$ and increases more 
$\tau_{\rm decay}$.
Thus, the models 1 give fast decay, models 2 intermediate decay and models 3 
slow 
decay, as is clear form Fig.~\ref{fig:cool_field_slow}.

The cooling influences the field decay by rising $\sigma$ till it becomes 
temperature
independent when dominated by impurity scattering. 
This happens during the photon cooling era as mentioned above, and happens 
earlier
for stiffer EOSs {\em given our choice of EOS independent $T_c$'s}. 
Consequently, we obtain that a stiffer EOS results in a slower field decay 
because
of its cooling behavior and also because of the larger length scale $l$.
Our choice of $T_c$ thus maximalize the effect of the EOS's stiffness on the 
field decay.

We  now turn our discussion to 
the analysis of GR effects upon magnetic field evolution.
For each model we simulated the field evolution with and without GR effects,
but note that GR effects on the star structure and cooling have always 
included. (Had we for instance, have turned 
off the GR effects on the star structure, the resulting models
 would make no sense at all
since the differences in the size, the central density etc, 
of the model would tender them unrealistic.) 
On the other hand, GR effects on 
on the cooling have already been discussed in the
literature long ago
(Nomoto \& Tsuruta 1987\cite{NT87}, Gudmundsson et al. 1983\cite{GPE83}).
The models with GR effects included are marked as `GR' in 
Fig.~\ref{fig:cool_field_slow}
and drawn with thick lines.
It is seen from those plots that the decay of the field is faster in NSs built
on a soft EOS than in the case of a medium or
stiff EOS but the decelerating GR effects are more pronounced.
This is most remarkable for long living fields:
the difference of the surface field strength for model 3, with the soft EOS, 
after $10^{10}$yrs of evolution is about two orders of magnitude. 
Comparing the field evolution for the soft and stiff EOS cases,
while the final surface field strength for model 3 in the stiff case is 
larger than 
those of the soft case by a factor of 300 when GR effects are neglected, that 
factor 
reduces to
about 7 when relativistic effects are taken into account correctly.

In the late photon cooling era, when most of the initial thermal energy of 
the NS
has been irradiated away, the Joule heating by the decay of the crustal field 
completely determines the cooling behavior.
Notice, however, that most of the magnetic energy has been dissipated earlier
when it had no effect on the cooling.
The amount of heat released in the process of Joule heating is determined by 
the strength of the field at that time and its decay rate.
Therefore, an initial field configuration as given by models of class 3 will 
result in a 
significant Joule heating while in models of class 1 the effect is practically
 nil.
The field decay of the class 2 models yields some noticeable Joule heating
only in case of the stiff EOS since its strength is large enough for that 
until about $10^8$yrs. 
For long periods Joule heating is especially effective in NSs with a stiff EOS:
their crust is quite thick, they keep a larger field and there is a lot of 
magnetic
energy to dissipate.
The effect of GR on the Joule heating is not spectacular but still significant
 in
the case of a soft EOS at the latest stages.
For the different EOSs it can result in stronger or weaker heating:
the rate of dissipation of magnetic energy is lower with GR but the magnetic
energy to be dissipated is larger due to the previous slower evolution and 
the net
effect can be an enhancement or a reduction of the heating.

Notice, finally, that at these late ages when the thermal evolution is entirely
controlled by the Joule heating the outer boundary condition, 
Eq.~\ref{equ:tb-te},
has no influence on the thermal evolution: this is due to the fact that the 
heating 
mechanism is temperature independent since $\sigma$ is totally controlled by 
impurity scattering.
The star's luminosity $L_*$ is thus simply given by the total heating rate
integrated over the whole star
\be
L_*^{\infty} = H \equiv \int q_h e^{2 \Phi} \; dV
\ee
since, given the low temperature at this times, $q_{\nu} \ll q_h$.
This is very fortunate for our study since the `$T_b - T_e$ relationship' is 
poorly
known at the low temperatures reached in this late phase of evolution.
It would of course not be the case if the heating mechanism were temperature 
dependent.

Notice finally that, in the case of magnetars, 
the Joule heating dominates the cooling from the very
beginning and thus it has a strong effect onto the field evolution 
(Geppert, Page, Colpi \& Zannias 1999\cite{GPCZ99}).

\subsection{Magneto-thermal evolution within the `fast' cooling scenario}

Enhanced neutrino emission is caused by the presence of exotic states of 
matter in 
the NS core. 
Kaon or pion condensation, hyperons, quarks as well as possible direct Urca 
processes 
enlarge the neutrino emissivity considerably. 
This results in a very fast cooling of the core, so that during the so called 
{\it isothermalization phase} the heat of the hotter crust is transported into
 the core. 
Depending on the EOS and the assumptions about pairing that phase can 
last of the order of 
1 to 100 yrs (see, e.g., Page 1998\cite{P98}). 
During this phase the surface temperature drops very rapidly and after this 
the crustal 
temperature 
is so low that the conductivity is almost completely controlled by 
electron--impurity 
collisions. 
Thus, for a period of time given by the impurity decay time scale 
$\tau_{\rm imp}$, 
the field suffers practically no decay. 
For $t > \tau_{\rm imp}$ the field decays according to a power law and, when 
it has diffused
down to the crust--core boundary, the decay becomes exponential. 
In class 3 models the power law like decay is missed because the field is 
already initially 
located close to the crust--core boundary so that it reaches this 
depth during $\tau_{\rm imp}$. 
Notice that $\tau_{\rm imp} \propto Q^{-1}$ and that it is also reduced  
with the softening of 
the 
EOS as an effect of the decreasing crustal thickness.
Generally, the field decay is slower than in the `standard' cooling scenario.

Due to the accelerated cooling, the Joule heating, even for the model 1 and 2 
fields, 
dominates the cooling earlier compared to the `standard' cooling scenario.
At ages above $\sim 10^7$ -- $ 10^8$ yrs, however, the star's temperature is 
comparable,  or even higher, to that
 predicted by the analogous models within the 
`standard' cooling scenario.
In class 2 and particularly class 3 models with the Stiff EOS, the stars' 
temperatures in this range of ages are noticeably higher, compared to the 
analogous 
`standard' cooling cases, since the magnetic field is higher and thus the 
joule 
heating more efficient.
After $10^{10}$yrs, however, the difference in the surface temperatures 
between the 
`standard' and the accelerated cooling scenario vanishes.

\subsection{Rotational evolution
             \label{sec:rot-evol}}

\begin{figure*}
\begin{center}
\resizebox{\hsize}{!}{\includegraphics{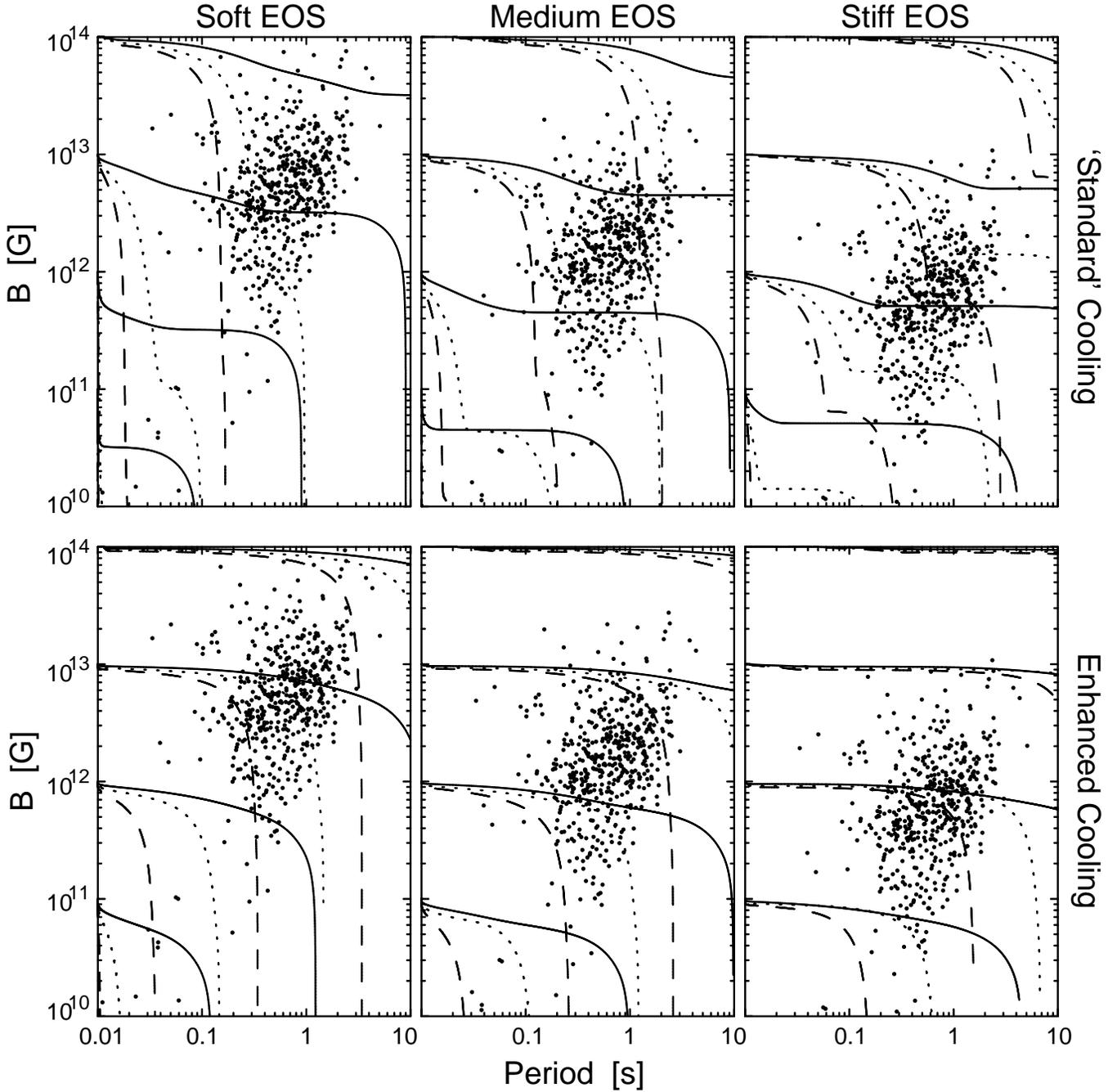}}
\caption{P-Pdot.
         Continuous lines correspond to initial current locations and
	 impurity contents of 
         $(\rho_0,Q_{\rm imp}) = (10^{14}$ g cm$^{-3}, 0.001)$,
         dotted lines to $(\rho_0,Q_{\rm imp}) = (10^{13}$ g cm$^{-3},0.01)$,
         and dashed lines to 
         $(\rho_0,Q_{\rm imp}) = (10^{12}$ g cm$^{-3},0.1)$.
	For each EOS, cooling scenario and initial field strength these three
        evolutionary tracks range from the most slowly decaying field 
        (continuous curve) to the fastest decaying field (dashed curve) and 
        thus encompass the whole range of predictions for the dynamical 
        evolution of a neutron star with a crustal field.}
\label{fig:B-P}
\end{center}
\end{figure*}

Given the temporal evolution of the magnetic field for the various EOSs and 
cooling 
scenarios, the rotational period of the NSs can be estimated 
by integrating the equation which relates the loss of rotational energy to the 
radiation of electromagnetic energy by magneto--dipole radiation
\be
P\,\dot P = \frac{2\pi^2}{3}\frac{R^6 \tilde{B}^2(t)}{c \, I},
\label{equ:dotP}
\ee
where $\tilde{B}$ is surface magnetic field at the magnetic pole,   
and $I$ is the star's moment of inertia.
A very accurately  relativistic expression for the moment of inertia,
 is given
by $I=0.21 \, e^{-2\Phi} \, M R^2 $ (Ravenhall \& Pethick 1994\cite{RP94}). 
However, this classical magneto--dipole radiation formula assumes flat space-time.
Ideally we would like to have the exact GR version of this equation.
To our knowledge, the GR version of it  remains to be calculated
and in the absence of the exact formula
we shall only use
an approximate expression.
We may recall that Eq.~\ref{equ:dotP} implicitly 
considers  the energy loss at the light cylinder
and then extrapolates  the field strength back to the star's surface assuming
the  flat space-time $1/r^3$ radial dependence of the field.
The full GR solution of a slowly rotating dipolar field, in vacuum, shows 
however
(Muslimov \& Tsygan 1990\cite{MT90}, 1992\cite{MT92}) that the actual
field at the star's magnetic pole is amplified by a factor
$f = -3y^2 [y \ln(1-y^{-1}) +\frac{1}{2}(2+y^{-1})]$,
where $y \equiv R/R_S$, compared to the flat
space-time one.
For instance, for our soft EOS star of 1.4~\Msun,
the resulting value of the amplification factor is  $f = 1.86$ while for the 
stiff one,
$f = 1.27$ (Page \& Sarmiento 1996\cite{PS96}).
In view of the Muslimov-Tsygan results, we will henceforth
 relate the surface magnetic field $B$ at  magnetic pole
as we numerically calculated, it to the corresponding
field $\tilde{B}$ used in Eq.~\ref{equ:dotP}
for the rotational evolution by $B = f \; \tilde{B}$.
For each of our three EOS (and a fixed stellar mass of 1.4 \Msun)
and the two cooling scenarios, we specify the initial field strength
and current depths $\rho_0$ and finally the impurity content $Q_{\rm imp}$.
The specification of those parameters yields a
unique evolution for  the surface $B$-field
and via Eq.~\ref{equ:dotP} the corresponding $P$. 
The resulting magneto-rotational evolution of isolated NSs is presented in 
Fig.~\ref{fig:B-P}.

We have also plotted, as dots, the PSRs data of the Princeton catalogue
(Taylor, Manchester, \& Lyne 1993\cite{TML93}):
their $B$ field is calculated with Eq.~\ref{equ:dotP}, using $P$ and $\dot P$ 
from 
that catalogue and the appropriate GR `$f$ factor' for each stellar model.
The  star's radius $R$ and moment of inertia $I$ are
correspond to values of the corresponding models having  mass of 1.4~\Msun.
The set of $B$--$P$ diagrams of Fig.~\ref{fig:B-P}
is the arena where our theoretical models confront the 
reality. 
There is a major effect on the data interpretation arising from varying the 
EOS:
we see that given the observed $P$--${\dot P}$ data
the ``observed'' surface $B$-field of pulsars differ almost one order of 
magnitude  
when we consider models where the equation of state  is varied
from the extreme soft one up to the extreme stiff one
(see also UK97\cite{UK97}).
This effect is a straightforward consequence of the fast growth of the 
magnetic 
field at small distance from the star which is moreover amplified by GR 
effects, combined with the much smaller stellar radius obtained by employing
the  soft EOS .
It should emphasized however,
that although a  complete and accurate model of pulsar 
spin-down will probably alter the simple Eq.~\ref{equ:dotP}
(expected at  least to modify the overall numerical coefficient) it 
certainly 
will not change the radial dependence of the field in the near zone.

A basic feature  emerging
from the Fig.(7), is the following:
a given model is viable if it manages to  maintain
a strong enough field for a long enough time such that its rotational 
period can increase
up to values compatible with the bulk of the observed pulsar.
It is also clear  that if a `standard' cooling scenario applies 
and irrespectively of the EOS, the field decays much faster than in the  case 
of the accelerated cooling:
this results in smaller saturation values of the rotational periods compared
to the accelerated cooling model, with the same initial field structure and 
the 
same EOS.

The evolutionary tracks offer the possibility
to check the whether a given EOS
can be naturally compatible with the observational
constraints without extreme ``fine tuning'' of the
initial parameters $\rho_0$ and $B_0$, of the impurity parameter $Q_{\rm imp}$ 
and its preference to standard versus enhanced cooling.
Thus for instance:
assuming a soft EOS, a comparison of the evolutionary tracks with the 
observed pulsar 
population strongly favors the enhanced cooling scenario.
Within the `standard' cooling scenario, only the strongest initial fields
$\sim 10^{14}$ G are acceptable.
In this case, even $B_0 \sim 10^{13}$ G requires very high $\rho_0$ and very 
low $Q_{\rm imp}$, definitely making the `standard' cooling scenario with a 
soft
EOS very unappealing.
Notice that the fast cooling scenario is very natural for a soft EOS since the
central densities reached in NSs are above ten times nuclear matter density.
Even if the pulsar fields (for a given $P$) were much lower than estimated we 
would 
still be in the same situation since the high field is required to be able to 
spin 
down the pulsars to the range of observed $P$'s before the field decays too 
much.
In the case of a medium EOS the requirements are much less stringent.
Within the `standard' cooling scenario, initial fields of the
order of 10$^{12}$ G require high initial depths $\rho_0$ and low pollution
in order for the evolutionary tracks to reach the bulk of the pulsar
population.
Finally, in the case of a stiff EOS, the field decay is slow enough that the 
whole pulsar population is
reachable with initial fields in the range 10$^{11}$ - 10$^{13}$ G
without basically any significant restrictions on $\rho_0$ and $Q_{\rm imp}$
in both the fast and the slow cooling scenarios.

One cannot overemphasize that the above analysis is
plagued by the intrinsic uncertainties of Eq.~\ref{equ:dotP}
and should be considered as only indicative.
Moreover, it is worth stressing here  that tracks starting with an initial 
field 
$B_0 = 10^{14}$G have to be considered with reserve since for such 
(and larger) 
field strengths the decay may not any longer described by a linear diffusion 
equation. 
The possible occurrence of a Hall--cascade 
(Goldreich \& Reissenegger 1992\cite{GR92}) 
would imply that the field evolution would deviate  significantly  
from  Eq.~\ref{equ:IndequF}, 
which is strictly valid in the limit that the magnetization parameter 
$\omega_B\tau < 1$ ($\omega_B$ being the Larmor frequency and $\tau$ 
the relaxation time of the electrons as given by Eq.~\ref{equ:tau}).

\section{Discussion and conclusions}

We studied in detail, and for a large variety of possible models, 
the magnetic, 
thermal and  rotational evolution of isolated NSs, assuming that their 
magnetic 
fields, and the currents supporting them, are confined to the stellar crust. 
Our calculations take into account, for the first time, all mutual effects of 
the thermal and magnetic evolution self consistently in a wholly GR formalism. 

\subsection{Comparison with the work of Urpin and Konenkov}

Urpin \& Konenkov\cite{UK97} (1997, UK97 thereafter) have presented the
most detailed study of the evolution of crustal neutron star magnetic field 
while
Miralles, Urpin \& Konenkov (1998\cite{MUK98}) completed the former work by 
the 
inclusion of Joule heating in  the thermal evolution.
It should be mentioned, however, that neither of these works incorporated GR 
effects 
on the $B$-field evolution.

Our field evolution results, without GR effects incorporated, are very close 
to those
of UK97 (their curves labeled 3, dashed lines, in their figure 2 correspond
to our models 2).
The main differences are seen in the early evolution of the field 
(our phase a):
this can be easily attributed to the fact that UK97 started the field 
evolution 
at an age of 100 yrs while we started it immediately at the star's birth. 
Since UK97 used isothermal stars for their modeling of the B-field evolution
they could not model the early hot, and non isothermal, phase properly.
In our calculations, this time difference allows the initial currents to 
rapidly
relax from their initial distribution (Eq.~\ref{equ:initial_F}, which 
{\em does not}
fulfill the outer condition) during a phase of low conductivity while the 
models
of UK97 must do it in conditions of higher conductivity, i.e., their relaxation
is much slower.
However, if one contemplates the scenario in which crustal magnetic fields
are generated by a thermomagnetic instability during this early hot phase
(Blandford et al. 1983\cite{BAH83}, 
Urpin, Levshakov \& Yakovlev 1986\cite{ULY86}, 
Wiebicke \& Geppert 1996\cite{WG96}) it is equally reasonable to start the 
B-field 
evolution by pure ohmic decay at the end of this phase as UK97 did.
The slight differences seen in the strength of the field at the onset of the
impurity scattering dominated phase (phase b) can be easily explained as due
to slight differences in the cooling histories, particularly differences 
in the time at which photon cooling takes over neutrino cooling
(the knee at ages around 10$^6$ -- 10$^7$ yrs) and also as a result of the
differences in the previous phase a.

Finally, the field evolution in the late phases, c and d, agree well with the
results of UK97 once we take into account the small differences in the
previous phases.

\subsection{General Relativistic effects}

While previous  studies considered the influence of GR effects into 
the star's structure and on its thermal evolutions 
(see e.g. Page 1989\cite{P89}), their incorporation into  
the magnetic field evolution had not been properly accounted for.
In the present paper we have been working
with the GR version of the diffusion equation, Eq.~\ref{equ:IndequF}, 
accompanied by proper boundary conditions, Eq.~\ref{equ:outbc_a},
derived in detail by GPZ00\cite{GPZ00}. 

The analysis of GPZ00\cite{GPZ00} clearly showed that most of the effect of 
GR on 
the field decay is due to the presence of  the red-shift factor $e^{\Phi}$ in
Eq.~\ref{equ:IndequF}. 
This can be intuitively understood by noticing that the red-shift factor 
relates
the proper time in each layer inside the star, i.e., the physical time in the 
layer
where the currents are located and decaying, to the time of an observer who is 
observing the decay, at infinity (i.e., the coordinate time).
As it can be easily seen from Eq.~\ref{equ:IndequF},
the effects of the red shift factor on the 
field decay could be approximated by a flat spacetime diffusion
equation running however on a slower time scale, given by some kind of
averaged red-shift.
This is clear from our 
Figures~\ref{fig:cool_field_slow} and \ref{fig:cool_field_fast} 
where the GR curves have the same shape as the non-GR ones but shifted to the 
right.

Comparing the field evolution when GR effects are included with
the evolution when they are neglected, 
we see that we obtain quite larger fields already in phase a, and 
in the late decay (phase c), we obtain much larger fields,
by up to two orders of magnitude for very compact stars.
GR effects in a field evolution can be viewed as almost equivalent to an  
evolution without GR effects but with currents initially located to higher
densities and, in the late decay phase, with lower impurity content.
This means that all constraints previously obtained about the location of the
currents and the impurity contents of the crust are significantly weakened when
GR effect are taken into account.

\subsection{Comparison with the work of Sengupta}

As far as the previously mentioned work of Sengupta (1998\cite{S98}) 
is concerned, although we are in qualitative agreement with his results,
we find many quantitative as well as interpretational differences.
For instance this author considered a soft EOS (similar to the one we used) 
and field
configurations with currents supposedly initially located at densities of 
$2 \times 10^{11}$ and $4 \times 10^{11}$ g cm$^{-3}$.
However, we obtain much higher densities at the depths at which he locates 
these 
layers:
using his values of $x \equiv r/R$ = 0.979 and 0.9834 we
find densities of the order of $6 \times 10^{13}$ and $2.6 \times 10^{13}$
g cm$^{-3}$ respectively.
Consequently, his field decay curves should be compared with our class 2 
models:
we see then a rough quantitative agreement with respect to the importance of 
the
GR effects with the significant difference that the late exponential decay 
(phase d) is absent in his GR models. 
As mentioned above, since the B-field evolutionary curves with and 
without GR effects should show approximately similar shapes
we deduce that the different behavior of the $B$ field at late times
obtained by Sengupta, must be due to numerical inaccuracies combined
with his neglect of appropriate boundary conditions at the stellar surface
as well as his employment of the Schwarzschild geometry.

\subsection{Constraining the neutron star EOS and cooling history
            within the crustal magnetic field hypothesis}

The analysis of the $B - P$ diagrams of \S~\ref{sec:rot-evol}
may be a tool to constrain the structure of matter at high density.
In a similar analysis, UK97 concluded that a stiff EOS
with `standard' cooling is the most promising model for understanding the 
observed 
pulsar population properties and that a medium EOS requires currents located 
at 
relatively high densities and low impurity contents.
These authors prefer the `standard' cooling scenario on the basis that it
implies an early decay of the field which may have an observational support
coming the fact the young pulsars with an associated supernova remnant have
stronger magnetic field than the bulk of the population.
Our results, with GR included, show that a medium EOS with `standard' cooling
is also compatible with the observed $P$ and $\dot{P}$ with much weaker 
constraints 
on the initial dipole strength $B_0$ and penetration density $\rho_0$ and on
the impurity content $Q_{\rm imp}$ than when GR effects are neglected.
A soft EOS with `standard' cooling needs very special conditions to
accommodate the observational data: large $B_0$, high $\rho_0$, and very low
$Q_{\rm imp}$.

But, notice that in case fast neutrino cooling is operating
any EOS could be compatible with the observed pulsar population 
with only weak constraints on $B_0$, $\rho_0$, and $Q_{\rm imp}$
as we argued in \S~4.4.
However, this cooling scenario is more physical in the case of a soft EOS and
probably incompatible with an EOS as stiff as our stiff EOS (for which the
central density of a 1.4 \Msun star does not even reach twice the nuclear
density).
In summary, it appears rather difficult within the crustal field hypothesis
alone to draw any strong conclusion about the EOS of neutron stars
and their cooling histories.

\subsection{The EOS and cooling history of neutron stars
            and the crustal magnetic field hypothesis}

There exist independent arguments in favor of a not too soft neutron star EOS,
none of them being, at the present time, compelling.
Interpretation of kilohertz quasi-periodic oscillations (KHz QPO's) in
several low mass X-ray binaries indicate these systems contain neutron stars
with masses around 2 \Msun (Klu\'zniak 1998\cite{K98}) which a very soft EOS
is not able to sustain. 
The strong gravitational light bending around a very compact star
would make it almost impossible for such a star to show any modulation in its
surface thermal emission, in contradistinction to some observations
(Page \& Sarmiento 1996\cite{PS96}).

With regard to the thermal evolution of neutron stars,
comparison of cooling models with current estimate of surface temperature of 
young
neutron stars shows no clear evidence of occurrence of fast cooling
(\"Ogelman 1995\cite{O95}; Page 1998\cite{P98}).
This means that if these neutron stars do contain some `exotic' phase of 
matter,
its strong neutrino emission must be quenched by pairing 
(Page \& Applegate 1992\cite{PA92}; Page 1998\cite{P98})
so that their thermal evolution is very close to the `standard' one.

These two lines of arguments are consistent with the conclusions
arising from the crustal field hypothesis, including GR effects: 
the EOS of neutron stars is away from the very soft regime and their thermal
evolution is close to the prediction of the `standard' model.

\subsection{Joule heating and detectability of \\ Old Isolated Neutron Stars}

Our study of the effect of the Joule heating produced by the decaying
currents gives results very similar to the ones of 
Miralles et al (1998\cite{MUK98}) 
and shows that the GR effects on the field evolution do not introduce any
important  change on the resulting late time thermal evolution.

Another consequence of the GR effects on the B-field evolution is that we
can predict significantly stronger field strength over a Hubble time.
This implies that Old Isolated Neutron Stars will be spinning very slowly
and are hence more likely to be able to accrete matter from the interstellar
medium. 
As a consequence this significantly increases the chances of detecting
them through their thermal radiation, either due to the Joule heating
or to the accretion.

\section{Conclusions}

The effect of GR on both the thermal and magnetic evolution is to slow it
down, mostly because the proper time inside the star runs more slowly than
the observer's time.
This effect is of course stronger the more compact the star is.
On the other hand, with increasing compactness the field decay, as any 
diffusion
process, is accelerated because of the reduction of the length scale.
The competition of these two opposite tendencies reduces the sensitivity of the
field evolution to the softness or stiffness of the EOS.
As a result it is difficult to draw conclusions about the nature of the
dense matter EOS from magnetic field evolution studies alone.
However when taking into account information from other approaches, as
the cooling history, gravitational lensing, kHz QPO's, a consistent
picture of neutron star structure and evolution is obtained, in which the
EOS is not too soft and the cooling history is close to the so called 
`standard'
cooling scenario.

Reversely, we can consider these results as an argument in favor of the
crustal magnetic field hypothesis.
It is seen to be compatible with the observed distribution of pulsars in the
$P$--$\dot{P}$ diagram without requiring fine tuning of the models' parameters.
Moreover, these model parameters are consistent with values deduced from
various other neutron star studies.

\appendix
\section{Induction Equation and Joule Heating Rate}

In this appendix, we  shall provide a few of the intermediate 
calculations leading to the derivation of the relativistic expression for
the Joule heating employed in the numerical computations.
We begin by recalling that the covariant form of Maxwell's equations are 
as follows (Misner et al. 1973\cite{MTW73}; Wald 1984\cite{W84}):
\be
{\nabla}^{\alpha}F_{\alpha\beta}=-{4{\pi}\over c}J_{\beta}
\label{equ:max_1}
\ee
\be
{\nabla}_{[{\alpha}}F_{{\beta\gamma}]}=0
\label{equ:max_2}
\ee
where $F_{\alpha\beta}=-F_{\beta\alpha}$, $J_{\alpha}$ and ${\nabla}$
are the coordinate components of the Maxwell tensor, the  conserved four 
current 
and the covariant derivative operator, respectively.
We may first recall that once  a solution  $F_{\alpha\beta}$ of 
Eqs.~\ref{equ:max_1}--\ref{equ:max_2} has been specified, 
an observer with four velocity  $U^{\alpha}~,~U^{\alpha}U_{\alpha}=-1$
measures electric and magnetic fields $(E~,~B)$  given respectively by:
\be
E_{\alpha}=F_{\alpha\beta}U^{\beta}~~,~~B_{\alpha}=-
{1\over 2}{{\epsilon}_{\alpha\beta}}^{\gamma\delta}
F_{\gamma\delta}U^{\beta}
\label{equ:E_B}
\ee
where ${\epsilon}_{\alpha\beta\gamma\delta}$ stands for the four-dimensional
Levi-Civita tensor density. 
It follows then easily from Eq.~\ref{equ:E_B} that an inversion  yields:
\be
{F_{{\alpha\beta}}=U_{\alpha}E_{\beta}-U_{\beta}E_{\alpha}+
{\epsilon}_{\alpha\beta\gamma\delta}U^{\gamma}B^{\delta}}
\label{equ:E_B-2}
\ee
Thorne et al. (1982\cite{TM82}, 1986\cite{TPM86}) have introduced an elegant 
version of curved spacetime electrodynamics, the absolute space formulation,
that is reminiscent of the familiar flat spacetime electrodynamics formulated 
in terms of the electric and magnetic fields.
This can be accomplished by working directly with the physical frame components
of the electric and magnetic fields $({\vec E}~,~{\vec B})$, as determined by
the static observers relative to their orthonormal frames.
Recalling that such observers have a four velocity field 
$U^{a}=e^{\Phi}{\delta}^{a}_{o}$, (see Eq.~\ref{equ:schwarzschild}) then 
it follows that Maxwell's Eqs.~\ref{equ:max_1}--\ref{equ:max_2}
can be written in the following equivalent form:
\be
{\nabla}{\cdot}{\vec E}=4{\pi}{\rho}~,~~~
{\nabla}{\cdot}{\vec B}=0
\label{equ:Max_1}
\ee
\be
{{\vec \nabla}}{\times}(Z{\vec B})=
{4{\pi}\over c}Z{\vec J}+{1\over c}{\partial {\vec E}\over 
{\partial t}}
\label{equ:Max_2}
\ee
\be{{\vec \nabla}}{\times}(Z{\vec E})=
-{1\over c}{\partial {\vec B}\over {\partial t}}
\label{equ:Max_3}
\ee
where in this appendix  $Z$ stands for the lapse function related to the red 
shift 
factor by $Z=e^{\Phi}$ and the $\nabla$ operators are formed out of the 
following 
three metric: 
\be
{ds^{2}}_{3}=\frac{dr^2}{1 - 2 G m/c^2r} +
 r^2 d \Omega ^2.
\label{equ:3D-metric}
\ee
by the rules of vector calculus as applied to the above orthogonal coordinate 
system. 
It follows then rather easily that in the absence of any convective motion 
Eqs.~\ref{equ:Max_1}--\ref{equ:Max_3}, combined with Ohm's law
${\vec J}={\sigma}{{\vec E}}$, and within the MHD approximation 
(i.e., neglecting the displacement current in Eq.~\ref{equ:Max_2}), 
yield the following induction equation governing the evolution of the neutron
star magnetic field:
\be
{1\over c}{\partial {\vec B}\over {\partial t}}+
{\vec \nabla}{\times}[{c\over 4{\pi}{\sigma}}
{\vec \nabla}{\times}(Z{\vec B})]=0,
\label{equ:IndequB}
\ee 
The above form of the induction equation
combined with Eq.~\ref{equ:3D-metric} as well the line element 
Eq.~\ref{equ:schwarzschild} with the metric functions corresponding to a
non singular perfect fluid solution of Einstein's equations yield 
Eq.~\ref{equ:IndequF} of the main text (for a detailed derivation see 
GPZ00\cite{GPZ00}). 

We now consider the GR expression for the Joule heating. 
To do so, we start from the covariant expression of
the energy momentum tensor for an arbitrary electromagnetic field
(Misner et al. 1973\cite{MTW73}; Wald 1984\cite{W84}) i.e.:
\be
T_{\mu \nu}={1\over 4{\pi}}\left[F_{\mu \gamma}{F_{\nu}}^{\gamma}-
      {1\over 4}g_{\mu \nu}F_{\alpha \beta}F^{\alpha \beta}\right].
\label{equ:Tmunu}
\ee
Using the representation of Eq.~\ref{equ:E_B-2}, one can easily show that the 
electromagnetic energy density ${\cal E}$ as seen by the static observers is 
given by:
\begin{displaymath}
T_{\mu \nu} U^{\mu} U^{\nu} = 
{1\over 8{\pi}} [E^{\alpha} E_{\alpha}+B^{\alpha}B_{\alpha}] =
\end{displaymath}
\be
\;\;\;\;\;\;\;\;\;\;\;\;\;\;\;\;\;\;\;\;
{1\over 8{\pi}} [\vec{E} \cdot \vec{E} +\vec{B} \cdot \vec{B}] \equiv {\cal E}
\label{equ:EB_energy}
\ee
where $\vec{E}$ and $\vec{B}$ stand for the physical components of the 
electric and 
magnetic fields, respectively, satisfying 
Eqs.~\ref{equ:Max_1}--\ref{equ:Max_3}.
Poynting's theorem now has the following form
\be
\frac{\partial \cal E}{\partial \tau} = 
- \nabla \cdot {\vec{\cal S}} + 2 \, {\vec g} \cdot {\vec{\cal S}} 
- {\vec j} \cdot {\vec E}
\label{equ:poynting}
\ee
where
\be
{\vec{\cal S}} \equiv \frac{c}{4 \pi} ({\vec E} \times {\vec B})
\ee
is the Poynting vector and
\be
{\vec g} \equiv - \frac{\nabla Z}{Z} = - \nabla \Phi
\ee
is the gravitational acceleration,
$d \tau \equiv Z \; dt$ being the proper time interval.
The second term in the r.h.s. of Eq.~\ref{equ:poynting} is a purely 
relativistic
effect which results from the inertia of (electromagnetic) energy
(see, e.g., Thorne et al 1986\cite{TPM86}).
The only dissipative term in Eq.~\ref{equ:poynting} is the last one,
i.e., the Joule heating term, using Ohm's law:
\be
Q^{(\rm GR)}_h=\frac{\vec j \vec j}{\sigma}
\label{equ:Relq}
\ee
By construction $Q^{(\rm GR)}_h$ represents energy per unit (proper) time and
(proper) volume.


\bigskip
{\bf Acknowledgments.} This work was supported by a binational grant
DFG (grant \#444 - MEX - 1131410) -
Co\-na\-cyt (grant \#E130.443),
Co\-na\-cyt (grant \#2127P - E9507),
UNAM - DGAPA (grant \#IN105495)
and Co\-ordi\-na\-ci\'on Cien\-t\'{\i}\-fi\-ca - UMSNH. 
D.P. and T.Z. are thankful to the
Astro\-phy\-si\-ka\-li\-sches Inst\-it\-ut Potsdam for its kind hospitality and
U.G. to the Ins\-ti\-tu\-to de As\-tro\-no\-m\'{\i}a of UNAM.


%

\end {document}